\documentclass[aps,pre,preprint,superscriptaddress]{revtex4-1}  
\usepackage{graphicx}
\usepackage{subcaption}
\usepackage{dcolumn}
\usepackage{amssymb}
\usepackage{amsmath}
\usepackage{newtxtext,newtxmath,lipsum}
\usepackage[shortlabels]{enumitem}

\begin{document}
	\title{Transmission of Positons in a Modified Noguchi Electrical Transmission Line}
	
	\author{N. Sinthuja}
	\affiliation{Department of Physics, Anna University, Chennai - 600 025, Tamilnadu, India}
	
	\author{M. Senthilvelan}
	\affiliation{Department of Nonlinear Dynamics, Bharathidasan University, Tiruchirappalli - 620 024, Tamilnadu, India}

    \author{K. Murali}
	\affiliation{Department of Physics, Anna University, Chennai - 600 025, Tamilnadu, India}
    
	\email{sinthukum14@gmail.com}
	
	\begin{abstract}
		\par  In previous studies, the propagation of localized pulses (solitons, rogue waves and breathers) in electrical transmission lines has been studied. In this work, we extend this study to explore the transmission of positon solutions or positons in the modified Noguchi electrical transmission line model. By converting the circuit equations into the nonlinear Schr\"odinger equation, we identify positons, a special type of solution with algebraic decay and oscillatory patterns. Unlike solitons, which are used for stable energy transmission, positons provide persistent energy localization and controlled spreading over long distances. We consider second-order and third-order positon solutions and examine their transmission behaviour in electrical lines. We show that over long times, the amplitude and width of both second- and third-order positons remain largely unaffected, indicating stable transmission. We also analyze what are the parameters affect the amplitude and localization of this kind of waves. Our investigations reveal that the amplitude and localization of positons are significantly influenced by the parameter $\epsilon$ that appear in the solution. Our findings have practical implications for improving energy transmission in electrical systems, where the management of wave localization and dispersion is crucial.
	\end{abstract}
	
	\maketitle

\section{Introduction}\label{sec1}

Nonlinear evolution equations are fundamental to a wide range of physical phenomena, including the modeling of wave dynamics in optics \cite{Solli2007optical}, the exploration of rogue waves in oceanography \cite{Dudley2019rogue}, the study of solitons in Bose-Einstein condensates (BECs) \cite{Luo2020solitons}, and the analysis of electrical transmission lines \cite{Ricketts2018electrical}. A study in these fields often relies on equations like Korteweg-de Vries (KdV), nonlinear Schr\"odinger (NLS) equation and sine-Gordon equation, which are known to exhibit multisoliton solutions, positons, negatons, and rational solutions \cite{Beutler1994what, Rasinariu1996negaton, Monisha2022higher, Zhou2023the, Akhmediev2006rogue}. Positons, a special type of solution, were first constructed by Matveev \cite{Matveev1992positon, Matveev1994a}. These solutions are oscillatory in nature, decay slowly, and act as long-range analogues of solitons \cite{Matveev1992positon, Matveev1994a}.

Nonlinear evolution equations can produce soliton solutions through the Lax pair approach, particularly when negative eigenvalues are employed in the Darboux transformation method. These solitons correspond to bound states of the Schr\"odinger operator associated with negative energy. In contrast, positive eigenvalues of the Lax pair equations lead to periodic solutions, which often lack substantial physical relevance. However, by carefully expanding these obtained periodic solutions at a chosen eigenvalue and incorporating them into the Wronskian framework, a new type of localized structure, termed positons, can be formed. These positons emerge due to positive spectral singularities embedded within the continuous spectrum. The evolution of positon solutions in the context of the KdV equation have been comprehensively studied in earlier research \cite{Maisch1995dynamic}. Positons are also considered potential models for shallow-water rogue waves, which are characterized by their destructive nature and extreme behavior. Beyond this, positon solutions have been observed in diverse areas, such as oceanographic studies \cite{Dubard2010on}, nonlinear optical systems, and ferromagnetic spin chain models \cite{Monisha2022higher}.


Positon solutions have been derived for various integrable systems, including the modified KdV equation \cite{Stahlhofen1992positon}, Toda lattice \cite{Stahlhofen1995positons} and sine-Gordon equation \cite{Beautler1993positon}. Smooth positons, which are non-singular solutions on vanishing backgrounds, have also been constructed for equations such as the generalized NLS equation \cite{Monisha2022higher}, the derivative NLS equation \cite{Song2019generating} and few other nonlinear evolution equations. Recently, breather-positon (B-P) solutions, representing positons on non-vanishing backgrounds, have gained attention due to their resemblance to rogue wave patterns. These solutions have been explored in systems such as the complex modified KdV equation \cite{Zhang2020novel}, Kundu-Eckhaus equation \cite{Qiu2019the}, and Sasa-Satsuma equation \cite{Guo2019darboux}.

The aforementioned works primarily focus on applications of positon solutions in optics, oceanography and plasma physics \cite{Dubard2010on, Monisha2022higher}. However, the impact of positon solutions have not been explored in the context of nonlinear electrical transmission lines. One notable nonlinear model is the modified Noguchi electrical transmission line, an advanced version of the classical transmission line model \cite{3,Aziz2020analytical,Kengne2022ginzeburg,Pelap2015dynamics,Marquie1994generation}. This model incorporates additional nonlinear and dispersive effects, enabling it to capture a wider range of wave dynamics often overlooked in simplified linear approximations. By studying the modified Noguchi line, one can better understand phenomena such as soliton propagation, rogue waves, and breathers, each with unique implications for signal stability, energy concentration, and system diagnostics. The transmission of these solutions, namely solitons, rogue waves, and breathers, has been studied in the context of the modified Noguchi electrical transmission line \cite{Djelah2023first,Kengne2017modelling,Kengne2019transmission,Duan2020super,Guy2018construction}. 

Theoretically, the dynamics of the modified Noguchi transmission line can be rigorously modeled using the NLS equation, one of the most studied equations in nonlinear wave theory. To derive the NLS equation from the modified Noguchi line, one has to start with the governing equations of the transmission line. By applying a multi-scale perturbation method, the system can be reduced to the NLS equation. In the the literature it has been analyzed how nonlinear waves (solitons, rogue waves, and breathers) arise and propagate in this electrical transmission line. However, no study has been made so far on (i) positons in electrical transmission lines, (ii) how they would propagate compared to solitons, rogue waves and breathers, and (iii) how higher-order positon solutions play a role in electrical transmission lines. In this work, we address all these questions.

We specifically consider positons for the following reasons: Among various nonlinear wave phenomena, positons stand out due to their distinct properties. Unlike solitons, which are single-peak waves that keep their shape and speed while traveling, positons have oscillating, multi-peak structures and decay slowly in space. In signal processing, disturbances can turn solitons into positons, which are less stable and more complex. Therefore, studying positons can help us to understand how and where they form in transmission lines and to find ways to control or convert them back into solitons in order to improve signal stability and energy flow. Studying positons in this way could lead to new ideas for advanced signal processing, energy harvesting, and fault detection.

In this work, we implement the derivation and realization of positons in electrical transmission lines, focusing on the modified Noguchi model as the basis for our analysis. By transforming the dynamics of this line into the NLS equation and solving for positons, we establish a theoretical framework that bridges nonlinear wave theory and electrical engineering. In this analysis, we consider second- and third-order positon solutions.  When time varies, the second- and third-order positons move forward without any change in their shape or amplitude for a certain period. Additionally, we shall demonstrate a small parameter $\epsilon$ plays a primary role in controlling the amplitude and localization of the second- and third-order positons. Through the numerical simulations and analytical insights, we demonstrate how positons can be generated, controlled, and utilized in the electrical transmission systems. This study not only advances our understanding of positons but also highlights their potential applications in modern electrical transmission technologies.

This paper is organized as follows: In Section $2$, we explain the model, the circuit equation, and how the NLS equation is derived from the circuit equation. In Section $3$, we describe the Darboux transformation for the standard NLS equation and present the second- and third-order positon solutions. In Section $4$, we examine the transmission of second- and third-order positons in the electrical transmission line. Finally, in Section $5$, we summarize our findings.
\section{Model and Circuit equation}
\begin{figure*}[!ht]
	\begin{center}
		\includegraphics[width=0.45\textwidth]{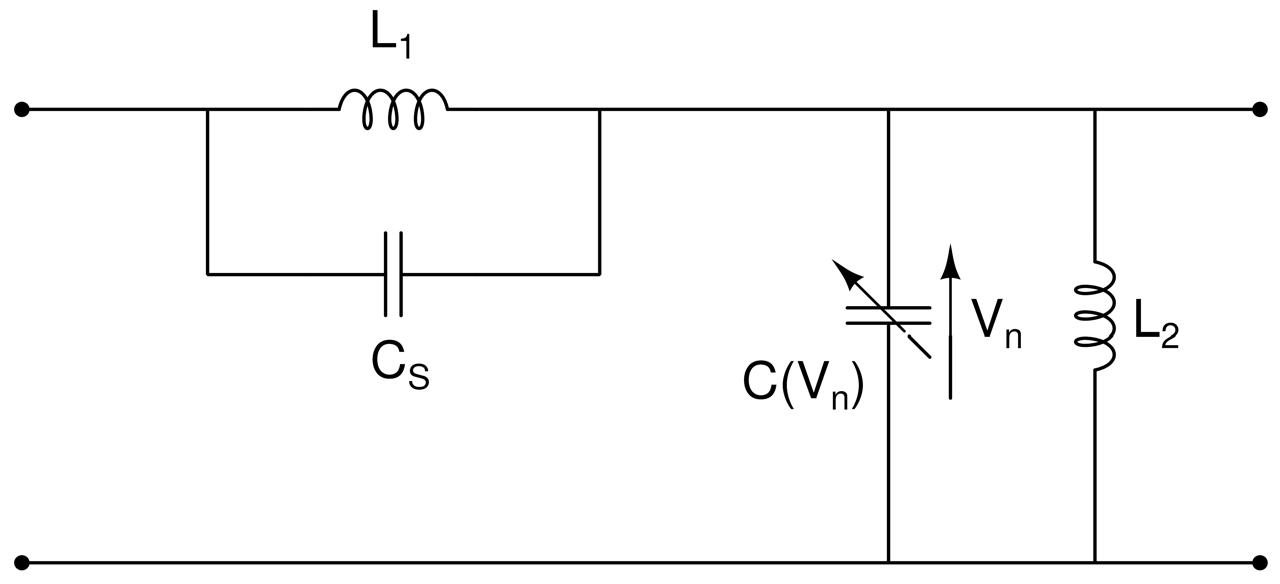}
	\end{center}
	\vspace{0.3cm}
	\caption{Schematic diagram of a discrete nonlinear electrical transmission line cell}
	\label{fig01}
\end{figure*}

The model under investigation is an one-dimensional electrical transmission line with dispersive and nonlinear properties. It consists of $N$ identical cells arranged in sequence. Each cell contains a linear inductor $L_1$ in parallel with a linear capacitor $C_S$, which enables the transmission and dispersion of electrical signals (see Fig. \ref{fig01}). This combination is connected in series with another parallel configuration containing a linear inductor $L_2$ and a nonlinear capacitor $C$, implemented with a reverse-biased diode. The nonlinear capacitance $C(V_n+V_b)$, which varies with the voltage $V_n$ across the $n$-th capacitor, follows a polynomial expansion  (for more details, please see Ref. \cite{Kengne2019transmission})
\begin{equation}
C(V_b+V_n)=\frac{dQ_n}{dV_n}=C_0(1-2\alpha V_n+3\beta V_n^2).
\label{eq1}
\end{equation}
where $C_0$ refers to the characteristic capacitance, while $\alpha$ and $\beta$ represent the nonlinear coefficients.

Applying Kirchhoff's laws, the wave propagation along the transmission line can be described by the set of discrete differential equations given below:
\begin{align}
\frac{d^2V_n}{dt^2}+u_0^2(2V_n-V_{n-1}&-V_{n+1})-\alpha\frac{d^2V_n^2}{dt^2}+\omega_0^2V_n+\beta\frac{d^2V_n^3}{dt^2}+\gamma \frac{d^2(2V_n-V_{n-1}-V_{n+1})}{dt^2}=0,
\label{eq2}
\end{align}
where $u_0=(L_1C_0)^{-\frac{1}{2}}$, $\omega_0=(L_2C_0)^{-\frac{1}{2}}$ and $\gamma=\frac{C_S}{C_0}$ represents the dispersive effect. The first term on the left hand side represents the acceleration (second time derivative) of the quantity $V_n$ at the $n$-th node. The second term models the interaction between neighboring nodes, where $(2V_n-V_{n-1}-V_{n+1})$ represents the difference in the state $V_n$ compared to its neighbors $V_{n-1}$ and $V_{n+1}$. The factor $u_0^2$ controls the strength of the coupling between neighbouring nodes. This term suggests that the system experiences a force (restoring force) proportional to the displacement difference between adjacent nodes, which is one of the characteristics of coupled oscillators or lattice systems. The third term introduces nonlinear damping or nonlinear stiffness. It is proportional to the second derivative of $V_n^2$ with respect to time, suggesting that the damping or restoring force depends on the square of the quantity $V_n$. This nonlinearity can model systems where the resistance or restoring force increases rapidly with displacement or velocity. The coefficient $\alpha$ determines the strength of this effect. The fourth term represents a linear restoring force acting on $V_n$, similar to the force in a harmonic oscillator (like a spring with Hooke's law). The coefficient $\omega_0^2$ corresponds to the natural frequency of the system. This term is similar to the restoring force in classical systems, where the force is proportional to the displacement. The fifth term involves a third-order nonlinearity. The second time derivative of $V_n^3$ suggests a nonlinear damping or restoring force that is dependent on the cube of $V_n$, representing a system where the damping force becomes more pronounced as $V_n$ increases. The coefficient $\beta$ controls the strength of this higher-order nonlinearity. The last term  is similar to the second term (coupling between neighbours) but it is modified by a higher-order time derivative. It represents nonlinear interactions between adjacent nodes, where the coupling force is influenced by the second time derivative of the displacement difference between neighbours. The term adds complexity to the coupling, indicating that the rate of change of the coupling itself might be nonlinearly dependent on the neighboring displacements. The coefficient $\gamma$ controls the strength of this interaction. Equation (\ref{eq2}) describes the dynamics of a discrete nonlinear system with coupled nodes indexed by $n$. This kind of system can exhibit interesting phenomena such as solitons, breather solutions, or rogue waves, depending on the specific parameter values \cite{Kengne2017modelling, Duan2020super, Djelah2023first, Kengne2019transmission,Guy2018construction}.

We introduce the slow variables $\zeta$ and $\tau$. For $\zeta$, we define $\zeta=\epsilon(n-v_g t)$ where $v_g$ represents the group velocity and  $n$ denotes the cell number. For $\tau$, we define $\tau=\epsilon^2 t$. In both cases $\epsilon$ denotes a small parameter. These variables capture the slow evolution of the wave's envelope as it propagates through the network. By adopting this scaling, we effectively smooth out rapid variations between cells, making the system easier to model and analyze as a continuous medium. We then assume that the solution of Eq. (\ref{eq2}) takes the following form \cite{Kengne2019transmission}:
\begin{equation}
V_n(t)=\epsilon\psi(\zeta,\tau)e^{i\theta}+\epsilon^2\psi_{10}(\zeta,\tau)+\epsilon^2 \psi_{20}(\zeta,\tau)e^{2i\theta}+\text{c.c}.
\label{eq3}
\end{equation}
Here, $\theta=(kn-\omega t)$ represents the phase that varies rapidly with both position and time, where $k$ is the wavenumber and $\omega$ is the angular frequency. The term 'c.c.' refers to the complex conjugate of the preceding expression. To account for the imbalance in the charge-voltage relationship described by Eq.(\ref{eq1}), the assumption in Eq.(\ref{eq3}) incorporates not only the fundamental term $\psi(\zeta,\tau)$, but also the DC component $\psi_{10}(\zeta,\tau)$ and the second-harmonic term $\psi_{20}(\zeta,\tau)$, thereby providing a more thorough characterization of the system.

Now we derive a simplified equation from the circuit Eq. (\ref{eq2}). The process begins by substituting the assumed form of the solution, given in Eq. (\ref{eq3}), into the original circuit Eq. (\ref{eq2}). After this substitution, we focus on isolating and keeping terms that are proportional to $\epsilon$, a small parameter often used to represent perturbations or nonlinear effects, as well as the exponential terms like $e^{i\theta}$, where $\theta$ is the rapidly varying phase. By retaining these specific terms, we obtain a more manageable equation that still captures the essential features of the system's dynamics.

For considering the terms involving $(\epsilon, e^{i\theta})$ and solving the resultant equation, we obtain the following linear dispersion relation and group velocity in the form
\begin{equation}
\omega=\sqrt{\frac{\omega_0^2+4u_0^2\sin^2(\frac{k}{2})}{1+4\gamma\sin^2(\frac{k}{2})}}, \quad
v_g=\frac{d\omega}{dk}=\frac{(u_0^2-\gamma\omega^2)\sin(k)}{\omega(1+4u_0^2\sin^2(\frac{k}{2}))}.
\label{eq5}
\end{equation}
Similarly, for $(\epsilon^4, e^{i0})$, we obtain the partial differential equation $\frac{\partial^2 \psi_{10}}{\partial \zeta^2}=\frac{2\alpha v_g^2}{v_g^2-u_0^2}\frac{\partial^2|\psi|^2}{\partial \zeta^2}$. Integrating this equation we obtain the solution (dc term) in the form
\begin{equation}
\psi_{10}(\zeta,\tau)=\frac{2\alpha v_g^2|\psi|^2}{v_g^2-u_0^2}+c_0(\tau)\zeta+c_1(\tau),
\label{eq8}
\end{equation}
where $c_0$ and $c_1$ are two arbitrary functions and they are real valued functions. 

When considering $(\epsilon^2, e^{2i\theta})$, we can obtain the expression for second harmonic term in the form
\begin{equation}
\psi_{20}(\zeta,\tau)=\frac{4\alpha\omega^2\psi^2}{4\omega^2+ 4(4\gamma\omega^2-u_0^2)\sin^2(k)-\omega_0^2}
\label{eq81}.
\end{equation}

While considering $(\epsilon^3, e^{i\theta})$, we obtain the following differential equation 
\begin{subequations}
	\label{eq9}
	\begin{equation}
	\label{eq91}
	i\frac{\partial \psi}{\partial \tau}+P\frac{\partial^2\psi}{\partial \zeta^2}+Q|\psi|^2\psi+\Gamma(\tau)\zeta\psi=0,
	\end{equation} 
	in which
	\begin{align}
	P&=\frac{1}{1+4\gamma\sin^2(\frac{k}{2})}\Bigg(-\frac{v_g^2}{2\omega}\left(1+4\gamma \sin^2(\frac{k}{2})\right)+\left(\frac{u_0^2}{2\omega}-\frac{\gamma \omega}{2}\right)\cos(k)-2\gamma v_g \sin(k)\Bigg),\\
	Q&=\frac{\omega}{1+4\gamma\sin^2(\frac{k}{2})}\Bigg(\frac{3\beta}{2}-\frac{2\alpha^2v_g^2}{v_g^2-u_0^2}-\frac{4\alpha^2\omega^2}{4\omega^2-\omega_0^2+4(4\gamma\omega^2-u_0^2)\sin^2(k)}\Bigg),\\
	\Gamma(\tau)&=-\frac{\alpha\omega}{1+4\gamma\sin^2(\frac{k}{2})}c_0(\tau).
	\label{eq10}
	\end{align}
\end{subequations}
Here, the dispersion coefficient $P$ describes the group velocity dispersion, whereas the self-modulation coefficient $Q$ measures the strength of the standard nonlinearity. 

In the literature, several works have appeared to transform the higher order NLS  type equations into more easily solvable forms (see Ref. \cite{r1,r2,r3,r4}). Based on one such transformation approach, equation (\ref{eq91}) can be transformed to the standard NLS equation. To do this, we assume the following transformation,
\begin{subequations}
	\label{eq11}
	\begin{equation}
	\psi(\zeta,\tau)=\rho \Psi(X,T)e^{i(\Lambda_1 \zeta+\Lambda_2)},
	\end{equation}
	with
	\begin{equation}
	X=\rho \sqrt{\frac{Q}{2P}}\zeta+V T,\quad T=\frac{1}{2}Q\rho^2\tau,\quad \frac{d\Lambda_1}{d\tau}-\Gamma(\tau)=0,\quad \frac{d\Lambda_2}{d\tau}+P\Lambda_1^2=0\quad V=\left( \frac{-4 \Lambda_{1} P}{Q \rho} \right) \sqrt{ \frac{Q}{2P} }.
	\label{ex1}
	\end{equation}
\end{subequations}
Substituting the above expressions (\ref{eq11}) into Eq. (\ref{eq9}) and rearranging, we end up at
\begin{equation}
i\frac{\partial \Psi}{\partial T}+\frac{\partial^2 \Psi}{\partial X^2}+2|\Psi|^2\Psi=0,
\label{eq12}
\end{equation}
the exact standard NLS equation. In Eq. (\ref{ex1}), we take $\Gamma(\tau)$ from Eq. (\ref{eq10}).  Here, we consider $c_0(\tau)$ as a real constant, where $c_0(\tau)=c_2$. Solving Eq. (\ref{ex1}) with $c_0(\tau)=c_2$ consequently, $\Lambda_1$ and $\Lambda_2$ take the following forms
\begin{align}
\Lambda_1&=-\frac{\alpha\omega}{1+4\gamma\sin^2(k/2)}c_2\tau+d_1,\\
\Lambda_2&=-P\Bigg(\frac{\alpha^2\omega^2c_2^2}{3(1+4\gamma\sin^2(k/2))^2}\tau^3-\frac{\alpha\omega c_2 d_1}{1+4\gamma\sin^2(k/2)}\tau^2+d_1^2\tau\Bigg)+d_2,
\end{align}
where $d_1$ and $d_2$ are integration constants. The NLS Eq. (\ref{eq12}) is integrable and it admits several kinds of localized solutions including soliton, breather and RWs.
\begin{figure*}[!ht]
	\begin{center}
		\begin{subfigure}{0.45\textwidth}
			\caption{}
			\includegraphics[width=\linewidth]{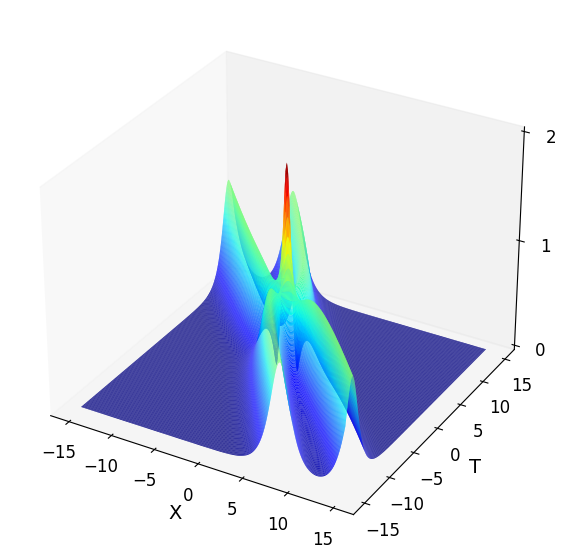}
		\end{subfigure}
		\begin{subfigure}{0.45\textwidth}
			\caption{}
			\includegraphics[width=\linewidth]{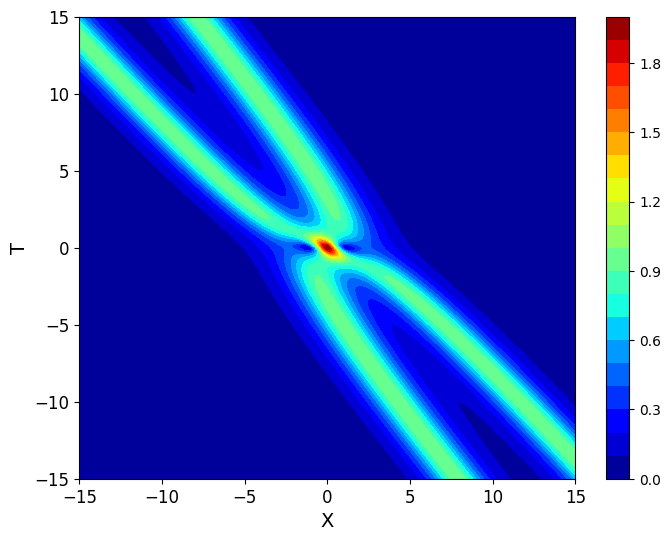}
		\end{subfigure}\\
		\begin{subfigure}{0.45\textwidth}
			\caption{}
			\includegraphics[width=\linewidth]{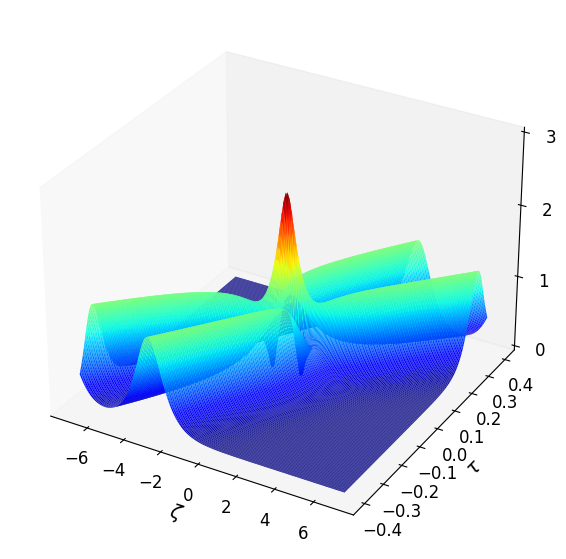}
		\end{subfigure}
		\begin{subfigure}{0.45\textwidth}
			\caption{}
			\includegraphics[width=\linewidth]{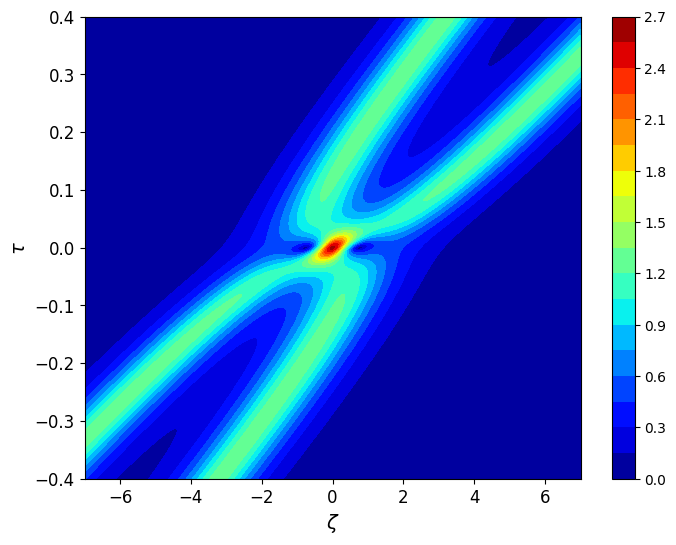}
		\end{subfigure}
	\end{center}
	\vspace{0.3cm}
	\caption{The surface plot illustrating the difference between the second-order positon (\ref{ep1}) and the second-order positon (\ref{ep1}) with experimental coordinates (\ref{eq11}) (i.e., (\ref{ep2})), in an electrical transmission line (\ref{eq2}) with the parameter $\lambda=0.2+0.5i$: (a) and (b) show the 3D plot and contour plot of respectively, while (c) and (d) show the 3D plot and contour plot of respectively.}
	\label{fig2}
\end{figure*} 

\section{Darboux transformation for the NLS Eq. (\ref{eq12})}
Equation (\ref{eq12}) can be represented in terms of the Lax pair equations. These Lax pair equations reproduce the original equation through a compatibility condition, as demonstrated in several papers \cite{Thulasidharan2024examining,Sinthuja2024rogue, Chen2018rogue}. The Lax pair of (\ref{eq12}) is \cite{Chen2018rogue}:
\begin{subequations}
	\label{ep0}
	\begin{align}
	\Phi_X=\begin{pmatrix}
	-i\lambda & \Psi \\
	-\bar{\Psi} & i\lambda
	\end{pmatrix} \Phi,\quad
	\Phi_T=\begin{pmatrix}
	-2i\lambda^2+i|\Psi|^2 & 2\lambda \Psi+i\Psi_X\\
	-2\lambda\bar{\Psi}+i\bar{\Psi}_X & 2i\lambda^2-i|\Psi|^2
	\end{pmatrix}\Phi
	\end{align}
\end{subequations}
where $\bar{\lambda}$ is the complex conjugation of a spectral parameter $\lambda$, $\Phi=(f_1,g_1)^T$ and $f_1$ and $g_1$ are nonzero solution of the Lax pair Eq. (\ref{ep0}). The $N$-fold DT formula for the NLS equation (\ref{eq11}) is \cite{Monisha2022higher,Su2018breather},
\begin{align}
\Psi^{[N]} (X,T) &= \Psi_0(X,T) - 2i \frac{|M_1^{[N]}(X,T)|}{|M_2^{[N]}(X,T)|}, \label{ep3}
\end{align}

\begin{align}
M_1^{[N]}(X,T) =
\begin{pmatrix}
\psi_1 & \phi_1 & \lambda_1 \psi_1 & \lambda_1 \phi_1 & \lambda_1^2 \psi_1 & \lambda_1^2 \phi_1 & \cdots & \lambda_1^{N-1} \psi_1 & \lambda_1^{N} \psi_1 \\
-\phi_1^{*} & \psi_1^{*} & -\lambda_1^* \phi_1^* & -\lambda_1^* \psi_1^* & -(\lambda_1^*)^2 \phi_1^* & -(\lambda_1^*)^2 \psi_1^* & \cdots & -(\lambda_1^*)^{N-1} \phi_1^* & -(\lambda_1^*)^{N} \phi_1^* \\
\psi_2 & \phi_2 & \lambda_2 \psi_2 & \lambda_2 \phi_2 & \lambda_2^2 \psi_2 & \lambda_2^2 \phi_2 & \cdots & \lambda_2^{N-1} \psi_2 & \lambda_2^{N} \psi_2 \\
\vdots & \vdots & \vdots & \vdots & \vdots & \vdots & \ddots & \vdots & \vdots \\
\psi_N & \phi_N & \lambda_N \psi_N & \lambda_N \phi_N & \lambda_N^2 \psi_N & \lambda_N^2 \phi_N & \cdots & \lambda_N^{N-1} \psi_N & \lambda_N^{N} \psi_N \\
\end{pmatrix}, \tag{12}
\end{align}

\begin{align}
M_2^{[N]}(X,T) =
\begin{pmatrix}
\psi_1 & \phi_1 & \lambda_1 \psi_1 & \lambda_1 \phi_1 & \lambda_1^2 \psi_1 & \lambda_1^2 \phi_1 & \cdots & \lambda_1^{N-1} \psi_1 & \lambda_1^{N-1} \phi_1 \\
-\phi_1^* & \psi_1^* & -\lambda_1^* \phi_1^* & -\lambda_1^* \psi_1^* & -(\lambda_1^*)^2 \phi_1^* & -(\lambda_1^*)^2 \psi_1^* & \cdots & -(\lambda_1^*)^{N-1} \phi_1^* & (\lambda_1^*)^{N-1} \psi_1^* \\
\psi_2 & \phi_2 & \lambda_2 \psi_2 & \lambda_2 \phi_2 & \lambda_2^2 \psi_2 & \lambda_2^2 \phi_2 & \cdots & \lambda_2^{N-1} \psi_2 & \lambda_2^{N-1} \phi_2 \\
\vdots & \vdots & \vdots & \vdots & \vdots & \vdots & \ddots & \vdots & \vdots \\
\psi_N & \phi_N & \lambda_N \psi_N & \lambda_N \phi_N & \lambda_N^2 \psi_N & \lambda_N^2 \phi_N & \cdots & \lambda_N^{N-1} \psi_N & \lambda_N^{N-1} \phi_N \\
\end{pmatrix}. \tag{13}
\end{align}
where $\Psi_0(X,T)$ is a seed solution. Using this DT formula, we can generate various solutions, including solitons, breathers, rogue waves and positon (or degenerate soliton) solutions \cite{Matveev1991darboux, Thulasidharan2024examining,Sinthuja2024rogue, Matveev1992positon}. For a zero seed solution $\Psi_0(X,T)$, the resulting $\Psi(X,T)$ provides one-soliton solution for the NLS equation at $\lambda=\lambda_1$. For a plane wave seed solution, the resulting $\Psi(X,T)$ provides a breather solution. By assuming and substituting a specific limiting process (depending on the equations) to the considered breather solution, we can obtain rogue wave solutions.
\begin{figure*}[!ht]
	\begin{center}
		\begin{subfigure}{0.45\textwidth}
			\caption{}
			\includegraphics[width=\linewidth]{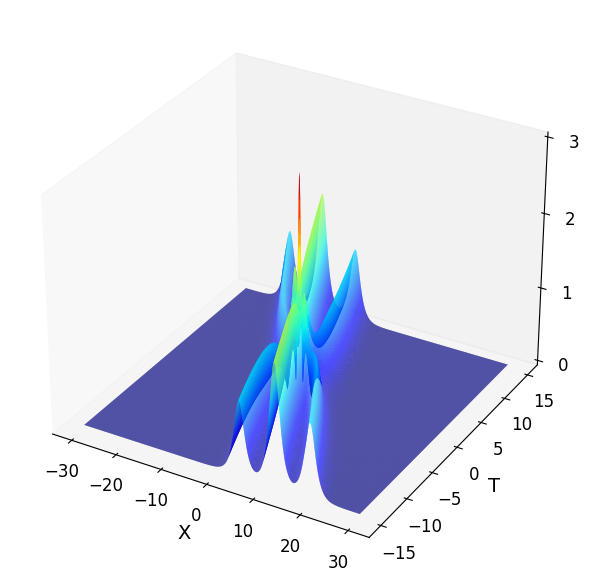}
		\end{subfigure}
		\begin{subfigure}{0.45\textwidth}
			\caption{}
			\includegraphics[width=\linewidth]{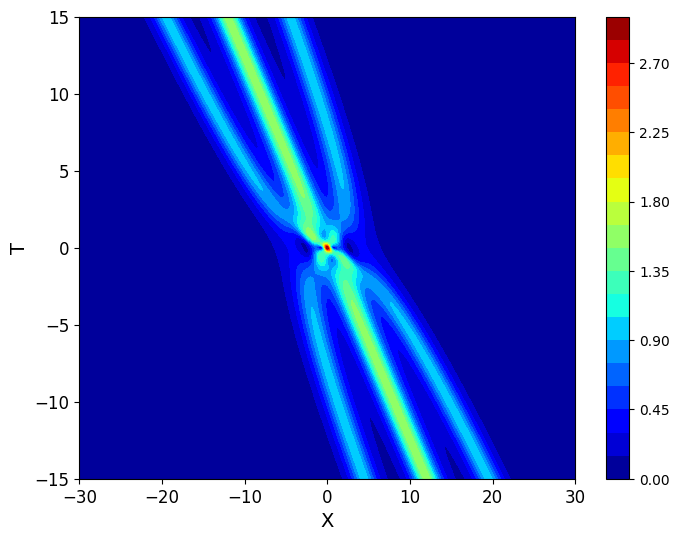}
		\end{subfigure}\\
		\begin{subfigure}{0.45\textwidth}
			\caption{}
			\includegraphics[width=\linewidth]{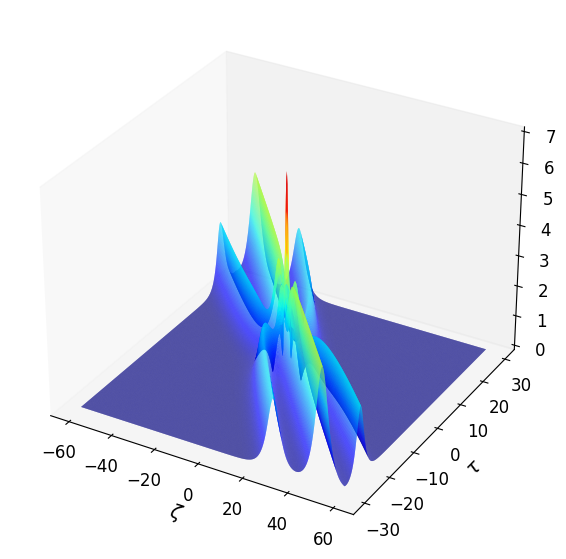}
		\end{subfigure}
		\begin{subfigure}{0.45\textwidth}
			\caption{}
			\includegraphics[width=\linewidth]{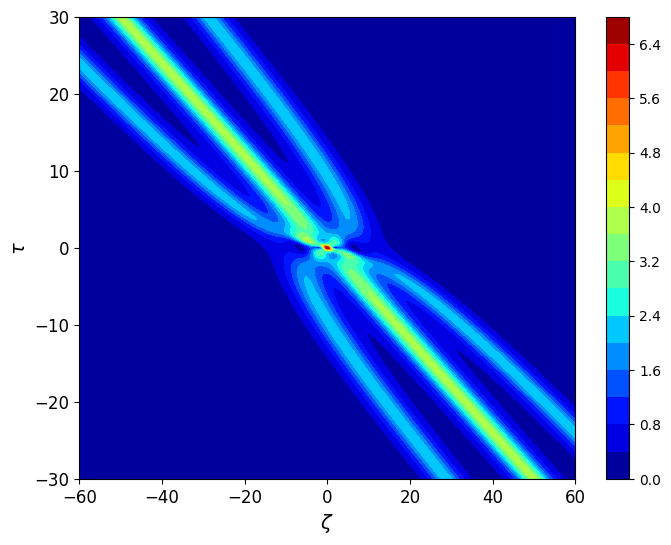}
		\end{subfigure}
	\end{center}
	\vspace{0.3cm}
	\caption{The surface plot illustrating the difference between the third-order positon (\ref{ep11}) and the third-order positon with experimental coordinates (\ref{eq11}) (the final form is not explicitly given), in an electrical transmission line (\ref{eq2}) with the parameter $\lambda=0.2+0.5i$: (a) and (b) show the 3D plot and contour plot of (\ref{eq12}), respectively, while (c) and (d) show the 3D plot and contour plot of (\ref{eq9}) with (\ref{eq11}), respectively.}
	\label{fig4}
\end{figure*} 

In the literature, the one-fold DT formula is used to generate one-soliton solution. Enforcing two-fold DT, we can derive second-order positon solutions at $\lambda_2=\lambda_1+\epsilon_1$, assuming $\Psi_0(X,T)$. The derivation of second-order positon solutions has already been presented in Ref. \cite{Monisha2022higher} and it takes the form (solution for the NLS Eq. (\ref{eq11}))
\begin{subequations}
	\label{ep1}
	\begin{align}
	\Psi^{[2]}(X, T) &= \frac{l_1(X, T)}{l_2(X, T)},
	\end{align}
	where $l_1(X, T)$ and $l_2(X, T)$ are give by
	\begin{align}
	l_1(X, T) &= 4 (\lambda_1 - \bar{\lambda}_1) \left( e^{( 2i X \lambda_1 + 4i T \lambda_1^2 )} \left( -4 T \lambda_1 (\lambda_1 - \bar{\lambda}_1) + X (\bar{\lambda}_1 - \lambda_1) - i \right) \right) \nonumber \\
	&\quad - 4 (\bar{\lambda}_1 - \lambda_1) \left( e^{( 2i \bar{\lambda}_1 X + 4i T \bar{\lambda}_1^2 )} \left( 4 T \bar{\lambda}_1 (\lambda_1 - \bar{\lambda}_1) + X (\lambda_1 - \bar{\lambda}_1) - i \right) \right),
	\end{align}
	\begin{align}
	l_2(X, T) &= e^{( 4i \bar{\lambda}_1 X + 8i T \bar{\lambda}_1^2 )} + e^{( 4i \lambda_1 X + 8i T \lambda_1^2 )} - 2 e^{( 2i X (\lambda_1 + \bar{\lambda}_1) + 4i T (\lambda_1^2 + \bar{\lambda}_1^2) )} \nonumber \\
	&\quad \times \left( 2 X^2 (\lambda_1 - \bar{\lambda}_1)^2 - 1 + 32 T^2 \lambda_1 \bar{\lambda}_1 (\lambda_1 - \bar{\lambda}_1)^2 + 8 T X (\lambda_1 - \bar{\lambda}_1)^2 (\lambda_1 + \bar{\lambda}_1) \right).
	\end{align}
\end{subequations}
From this solution we can get the solution of Eq. (\ref{eq9}) in the form
\begin{subequations}
	\label{ep2}
	\begin{align}
	\psi^{[2]}(\zeta, \tau) &= \rho\frac{l_1(\zeta, \tau)}{l_2(\zeta, \tau)}e^{i(\Lambda_1 \zeta+\Lambda_2)},
	\end{align}
	where $l_1(\zeta, \tau)$ and $l_2(\zeta, \tau)$ are give by
	\begin{align}
	l_1(X, T)& = 4 (\lambda_1 - \bar{\lambda}_1) \left[ e^{\left( 2i \rho \sqrt{\frac{Q}{2P}} \zeta \lambda_1 + 2i Q \rho^2 \tau \lambda_1^2 \right)} \left( -2 Q \rho^2 \tau \lambda_1 (\lambda_1 - \bar{\lambda}_1) + \rho \sqrt{\frac{Q}{2P}} \zeta (\bar{\lambda}_1 - \lambda_1) - i \right) \right]\notag\\
	&- 4 (\bar{\lambda}_1 - \lambda_1) \left[ e^{\left( 2i \rho \sqrt{\frac{Q}{2P}} \zeta \bar{\lambda}_1 + 2i Q \rho^2 \tau \bar{\lambda}_1^2 \right)} \left( 2 Q \rho^2 \tau \bar{\lambda}_1 (\lambda_1 - \bar{\lambda}_1) + \rho \sqrt{\frac{Q}{2P}} \zeta (\lambda_1 - \bar{\lambda}_1) - i \right) \right],\\
	l_2(X, T) &= e^{( 4i \bar{\lambda}_1 \rho \sqrt{\frac{Q}{2P}} \zeta + 4i Q\rho^2\tau \bar{\lambda}_1^2 )} + e^{( 4i \lambda_1 \rho \sqrt{\frac{Q}{2P}} \zeta + 4i Q\rho^2\tau \lambda_1^2 )} - 2 e^{( 2i \rho \sqrt{\frac{Q}{2P}} \zeta (\lambda_1 + \bar{\lambda}_1) + 2i Q\rho^2\tau (\lambda_1^2 + \bar{\lambda}_1^2) )} \nonumber \\
	&\times \left(  \rho^2 \frac{Q}{P} \zeta^2 (\lambda_1 - \bar{\lambda}_1)^2 - 1 + 8 Q^2\rho^4\tau^2 \lambda_1 \bar{\lambda}_1 (\lambda_1 - \bar{\lambda}_1)^2 + 4 \rho^3 Q \sqrt{\frac{Q}{2P}} \zeta\tau (\lambda_1 - \bar{\lambda}_1)^2 (\lambda_1 + \bar{\lambda}_1) \right).
	\end{align} 
\end{subequations}
Here $Q$, $P$ are given in Eq. (\ref{eq9}).

The surface plot of the second-order positon solution of Eq. (\ref{eq12}) and Eq. (\ref{eq9}) are given in Fig. \ref{fig2} for $\lambda=0.2+0.5i$. The first row of figures shows the 3D plot and the contour plot of $|\Psi^{[2]}(X,T)|$ for Eq. (\ref{eq12}). These are shown in Figs. \ref{fig2}(a) and \ref{fig2}(b), respectively. From both the 3D and contour plots, we observe that the maximum amplitude of second-order positon is approximately $2$ at the origin $(x,t)=(0,0)$. Similarly, Figs. \ref{fig2}(c) and \ref{fig2}(d) are plotted for $|\psi^{[2]}(\zeta,\tau)|$ of Eq. (\ref{eq9}) for the same eigenvalue. However, these solutions involve the experimental coordinates $\omega$, $v_g$, $P$ and $Q$, which we derived in the previous section. Using these experimental coordinates, we can fix $L_1=220 mH$, $L_2=470mH$, $C_0=370 \mu H$, $C_S=240.5 \mu H$, $\alpha=-0.5$, $\beta=-0.1445$ and $c_2=0.01$. Other integration constants $c_2,d_1$ and $d_2$ are considered as zero or otherwise cancel when taking the absolute value of the solution $|\psi^{[2]}(\zeta,\tau)|$. From Figs. \ref{fig2}(c) and \ref{fig2}(d), we notice that the orientation and width of the positon changes, and the positon position (that is the axis ranges) also shifts. Additionally, the maximum amplitude of the positon at $(0,0)$ is approximately $2.7$. These changes occur due to the influence of the experimental coordinates.

The third-order positon solution for the NLS Eq. (\ref{eq11}) reads \cite{Monisha2022higher}
\begin{subequations}
	\label{ep11}
	\begin{align}
	\Psi^{[3]}(X, T) &= \frac{s_1(X, T)}{s_2(X, T)},
	\end{align}
	\begin{align}
	s_1(X, T) = &\ 2i (\lambda - \bar{\lambda}) \Bigg( e^{4i\lambda (X + 2T\lambda)} \Big( -3 + 2X \Big(3i + 8T\lambda (\lambda - \bar{\lambda})\Big)(\lambda - \bar{\lambda}) + 2X^2 (\lambda - \bar{\lambda})^2 \nonumber\\&+ 32T^2 \lambda^2 (\lambda - \bar{\lambda})^2 + 4iT \Big(7\lambda^2 - 8\lambda\bar{\lambda} + \bar{\lambda}^2\Big)\Big) + e^{4i\bar{\lambda} (X + 2T\bar{\lambda})} \Big( -3 + 2X^2 (\lambda - \bar{\lambda})^2 \nonumber\\&+ 32T^2 (\lambda - \bar{\lambda})^2 \bar{\lambda}^2 + 2X (\lambda - \bar{\lambda})\Big(-3i + 8T (\lambda - \bar{\lambda}) \bar{\lambda}\Big) + 4iT \Big(\lambda^2 - 8\lambda\bar{\lambda} + 7\bar{\lambda}^2\Big)\Big) \nonumber\\&+ 2e^{2i X (\lambda + \bar{\lambda}) + 4i T (\lambda^2 + \bar{\lambda}^2)} \Big( -3 - 4X^2 \lambda^2 + 2X^4 \lambda^4 + 8X^2 \lambda \bar{\lambda}- 8X^4 \lambda^3 \bar{\lambda} - 4X^2 \bar{\lambda}^2 \nonumber\\&+ 12X^4 \lambda^2 \bar{\lambda}^2 + 512T^4 \lambda^2 (\lambda - \bar{\lambda})^4 \bar{\lambda}^2 - 8X^4 \lambda \bar{\lambda}^3 + 2X^4 \bar{\lambda}^4 + 16T (\lambda - \bar{\lambda})^2 \nonumber\\&\times\Big(i - i X^2 (\lambda - \bar{\lambda})^2 - X (\lambda + \bar{\lambda}) + X^3 (\lambda - \bar{\lambda})^2 (\lambda + \bar{\lambda})\Big)+ 64T^3 (\lambda - \bar{\lambda})^4 (\lambda + \bar{\lambda}) \nonumber\\&\times\Big(-i \bar{\lambda} + \lambda (-i + 4X \bar{\lambda})\Big) + 8T^2 (\lambda - \bar{\lambda})^2 \Big( 4X^2 \lambda^4 + 8X \lambda^3 (-i + X \bar{\lambda}) + \lambda^2 \Big(-1 \nonumber\\
	&+ 8i X \bar{\lambda} - 24X^2 \bar{\lambda}^2\Big) + 2\lambda \bar{\lambda} \Big(-3 + 4i X \bar{\lambda} + 4X^2 \bar{\lambda}^2\Big) + \bar{\lambda}^2 \Big(-1 - 8i X \bar{\lambda} + 4X^2 \bar{\lambda}^2\Big)\Big)\Big) \Bigg)
	\end{align}
	\begin{align}
	s_2(X, T) = &
	e^{6i\lambda (X + 2 T \lambda)} + e^{6i\bar{\lambda} (X + 2 T \bar{\lambda})} + 
	e^{2i (X (2 \lambda + \bar{\lambda}) + 2 T (2 \lambda^2 + \bar{\lambda}^2))} \bigg( 
	3 + 4 X^4 (\lambda - \bar{\lambda})^4\nonumber \\
	& + 1024 T^4 \lambda^2 (\lambda - \bar{\lambda})^4 \bar{\lambda}^2  + 48 T^2 (\lambda - \bar{\lambda})^2 (\lambda^2 - 6 \lambda \bar{\lambda} + \bar{\lambda}^2) 
	- 128i T^3 (\lambda - \bar{\lambda})^3\nonumber \\
	&\times  (\lambda^3 - 3 \lambda^2 \bar{\lambda} - 
	3 \lambda \bar{\lambda}^2 + \bar{\lambda}^3)  + 8 X^3 (\lambda - \bar{\lambda})^3\big(i + 
	4 T (\lambda^2 - \bar{\lambda}^2)\big) + 4 X^2 (\lambda - \bar{\lambda})^2 \nonumber \\
	&\times \big(-3 + 
	12i T (\lambda^2 - \bar{\lambda}^2)  + 16 T^2 (\lambda - \bar{\lambda})^2 (\lambda^2 + 
	4 \lambda \bar{\lambda} + \bar{\lambda}^2) \big) 
	+ 16 T X (\lambda - \bar{\lambda})^2\nonumber \\
	& \times \bigg( -3 \bar{\lambda} + 
	32 T^2 \lambda^4 \bar{\lambda} - 32 T^2 \lambda^3 \bar{\lambda}^2 - 8 T \lambda^2 \bar{\lambda} (-3i + 
	4 T \bar{\lambda}^2) + \lambda \big(-3 - 24i T \bar{\lambda}^2 \nonumber \\
	& + 
	32 T^2 \bar{\lambda}^4\big) \bigg) \bigg) + e^{2i (X (\lambda + 2 \bar{\lambda}) + 2 T (\lambda^2 + 2 \bar{\lambda}^2))} \bigg( 
	3 + 4 X^4 (\lambda - \bar{\lambda})^4 + 1024 T^4 \lambda^2 (\lambda - \bar{\lambda})^4 \bar{\lambda}^2  \nonumber \\
	& + 48 T^2 (\lambda - \bar{\lambda})^2 (\lambda^2 - 6 \lambda \bar{\lambda} + \bar{\lambda}^2) 
	+ 128i T^3 (\lambda - \bar{\lambda})^3 (\lambda^3 - 3 \lambda^2 \bar{\lambda} - 
	3 \lambda \bar{\lambda}^2 + \bar{\lambda}^3) \nonumber \\
	&+ 8 X^3 (\lambda - \bar{\lambda})^3 \big(-i  + 
	4 T (\lambda^2 - \bar{\lambda}^2)\big) + 4 X^2 (\lambda - \bar{\lambda})^2 \big(-3 - 
	12i T (\lambda^2 - \bar{\lambda}^2)\nonumber \\
	& + 16 T^2 (\lambda - \bar{\lambda})^2 (\lambda^2 + 
	4 \lambda \bar{\lambda} + \bar{\lambda}^2) \big) 
	+ 16 T X (\lambda - \bar{\lambda})^2  \bigg( -3 \bar{\lambda} + 
	32 T^2 \lambda^4 \bar{\lambda} \nonumber \\
	& - 32 T^2 \lambda^3 \bar{\lambda}^2 - 8 T \lambda^2 \bar{\lambda} (3i + 
	4 T \bar{\lambda}^2) + \lambda \big(-3 + 24i T \bar{\lambda}^2 + 
	32 T^2 \bar{\lambda}^4\big) \bigg) \bigg)
	\end{align}
\end{subequations}

Figure \ref{fig4} presents the surface plot of the third-order positon solution derived from Eq. (\ref{eq12}) and Eq. (\ref{eq9}) with $\lambda=0.2+0.5i$. The 3D representation of $|\Psi^{[3]}(X,T)|$ is depicted in Fig. \ref{fig4}(a), while its corresponding contour plot is shown in Fig. \ref{fig4}(b). Both the visualizations confirm that the maximum amplitude of the third-order positon reaches approximately $3$ at the origin, $(X,T)=(0,0)$. Similarly, Figures \ref{fig4}(c) and \ref{fig4}(d) illustrate $\Psi$, the solution of Eq. (\ref{eq9}) for the same eigenvalue. These solutions incorporated with the experimental parameters $\omega$, $v_g$, $P$, and $Q$, which were previously defined. The parameter values are chosen as $L_1=220 mH$, $L_2=470mH$, $C_0=370 \mu H$, $C_S=240.5 \mu H$, $\alpha=-0.5$, $\beta=-0.1445$ and $c_2=0.01$. Other integration constants, such as $d_1$, and $d_2$, are assumed to be zero or are canceled when calculating $|\psi^{[3]}(x,t)|$. From Figs. \ref{fig4}(c) and \ref{fig4}(d), it is observed that the orientation and width of the positon changes, along with a noticeable shift in its position (axis ranges). Additionally, the maximum amplitude at $(\zeta,\tau)=(0,0)$ increases to approximately 7. These variations are attributed to the influence of the experimental parameters.

\section{Transmission of positons in the modified electrical transmission line}
Until now, we have analyzed solutions using experimental coordinates. To further investigate the behavior of second order positons in the electrical transmission line, we consider the second-order positon solution of the circuit equation (\ref{eq2}), which is expressed in the form given in Eq. (\ref{eq3}) and is given by:
\begin{equation}
V^{[2]}_n(t)=Re(\epsilon\psi^{[2]}(\zeta,\tau)e^{i\theta}+\epsilon^2\psi_{10}(\zeta,\tau)+\epsilon^2 \psi_{20}(\zeta,\tau)e^{2i\theta}+c.c.),
\label{ep31}
\end{equation}
where $\psi^{[2]}(\zeta,\tau)$, $\psi_{10}(\zeta,\tau)$ and $\psi_{20}(\zeta,\tau)$ are given in Eqs. (\ref{ep2}), (\ref{eq8}) and (\ref{eq81}) respectively.

\begin{figure*}[!ht]
	\begin{center}
		\begin{subfigure}{0.45\textwidth}
			\caption{}
			\includegraphics[width=\linewidth]{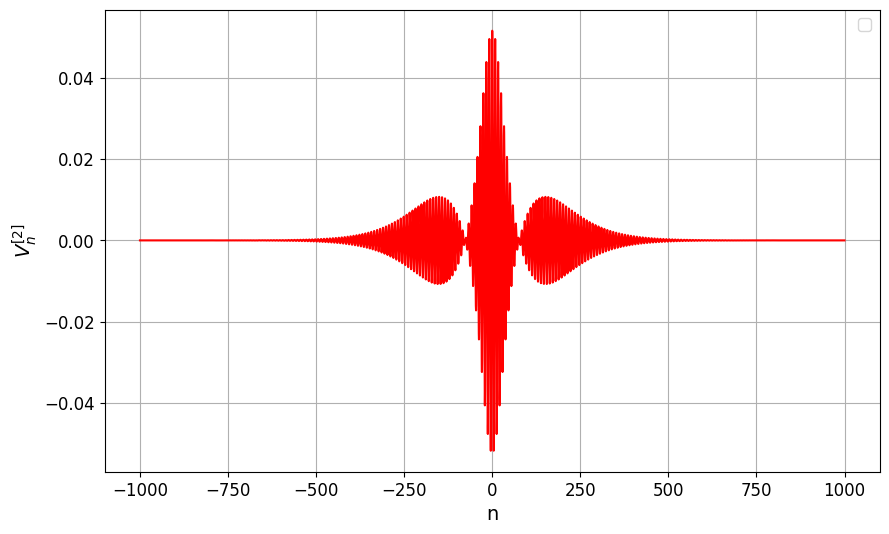}
		\end{subfigure}
		\begin{subfigure}{0.45\textwidth}
			\caption{}
			\includegraphics[width=\linewidth]{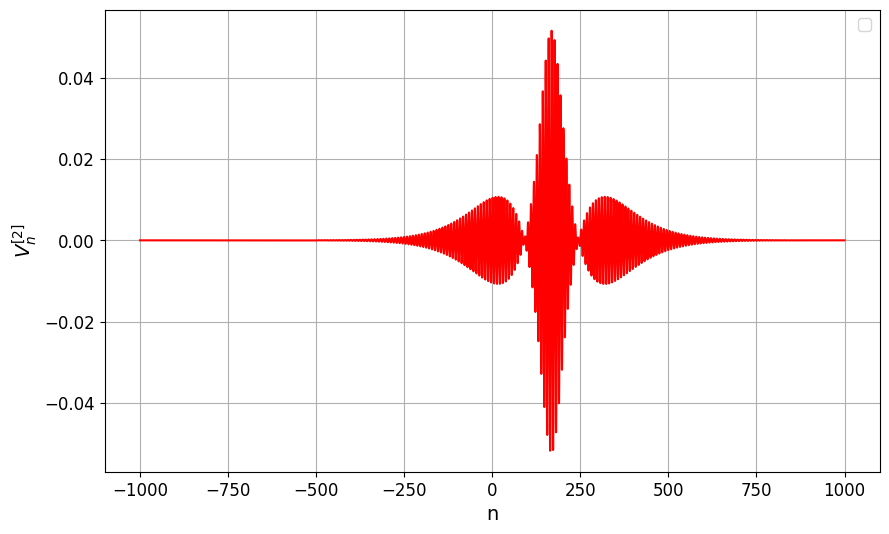}
		\end{subfigure}\\
		\begin{subfigure}{0.45\textwidth}
			\caption{}
			\includegraphics[width=\linewidth]{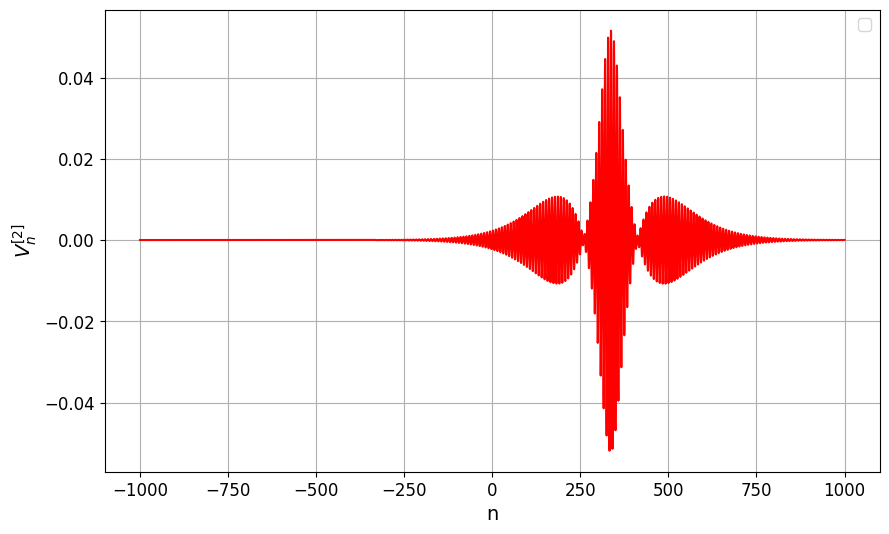}
		\end{subfigure}
		\begin{subfigure}{0.45\textwidth}
			\caption{}
			\includegraphics[width=\linewidth]{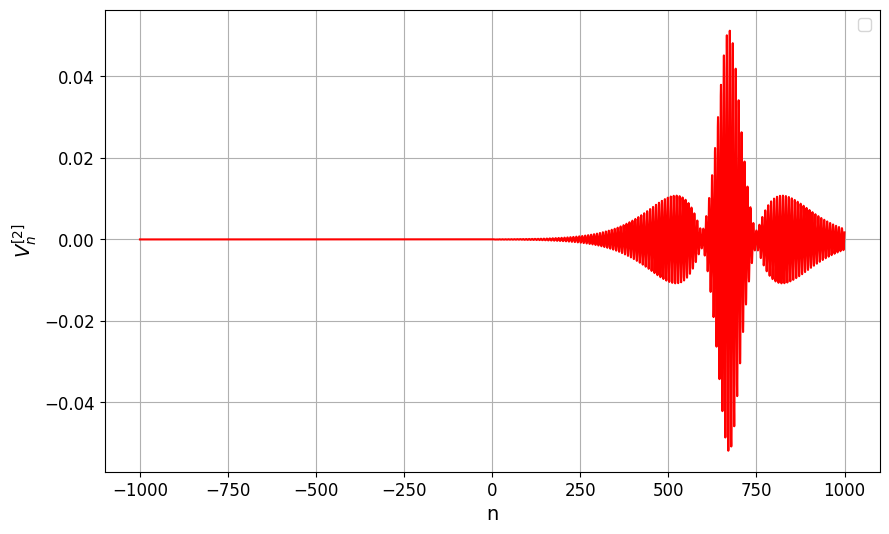}
		\end{subfigure}
	\end{center}
	\vspace{0.3cm}
	\caption{Transmission of second-order positon in an electrical transmission line (\ref{eq2}) for $\lambda=0.2+0.5i$ and different times: (a)  $t=0$; (b) $t=5$; (c) $t=10$; (d) $t=20$.}
	\label{fig3}
\end{figure*}

Figure \ref{fig3} demonstrates the transmission of a second order positon in the modified electrical transmission line (\ref{eq2}) using the same experimental coordinates as described in Fig. \ref{fig2} for $\lambda=0.2+0.5i$. To begin, we only change the value of time ($t$) and study how the positon travels as a function of time $t$. Figure \ref{fig3}(a) is obtained for $t=0$, where the positon occurs at the origin. As we increase the value of $t$ to $5$ and $10$, the positon begins to move forward, as shown in Figs. \ref{fig3}(b) and \ref{fig3}(c). Further increasing the time to $t=20$, the positon reaches near $n=1000$, where $n$ represents the cell number. From this, we conclude that as time increases, the positon propagates continuously through the network, specifically moving into a cell without any change in its shape.

\begin{figure*}[!ht]
	\begin{center}
		\begin{subfigure}{0.45\textwidth}
			\caption{}
			\includegraphics[width=\linewidth]{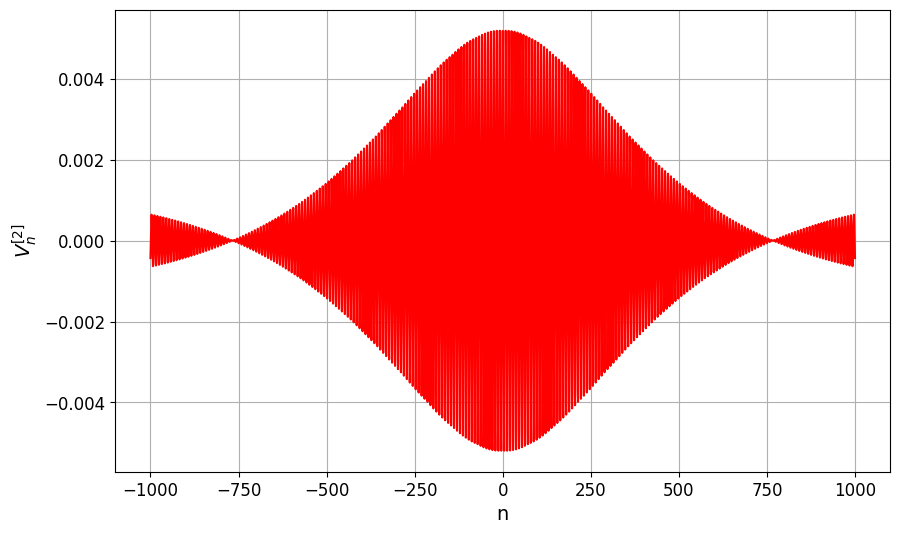}
		\end{subfigure}
		\begin{subfigure}{0.45\textwidth}
			\caption{}
			\includegraphics[width=\linewidth]{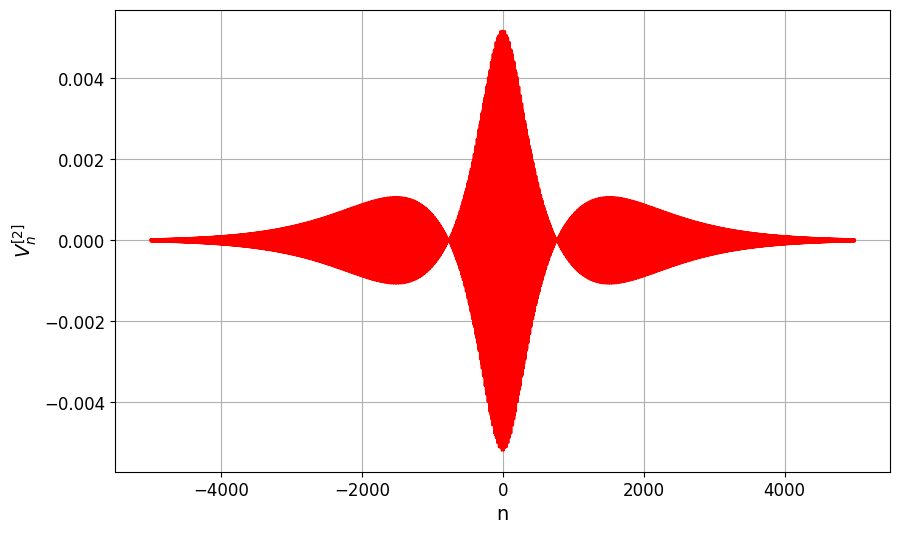}
		\end{subfigure}\\
		\begin{subfigure}{0.45\textwidth}
			\caption{}
			\includegraphics[width=\linewidth]{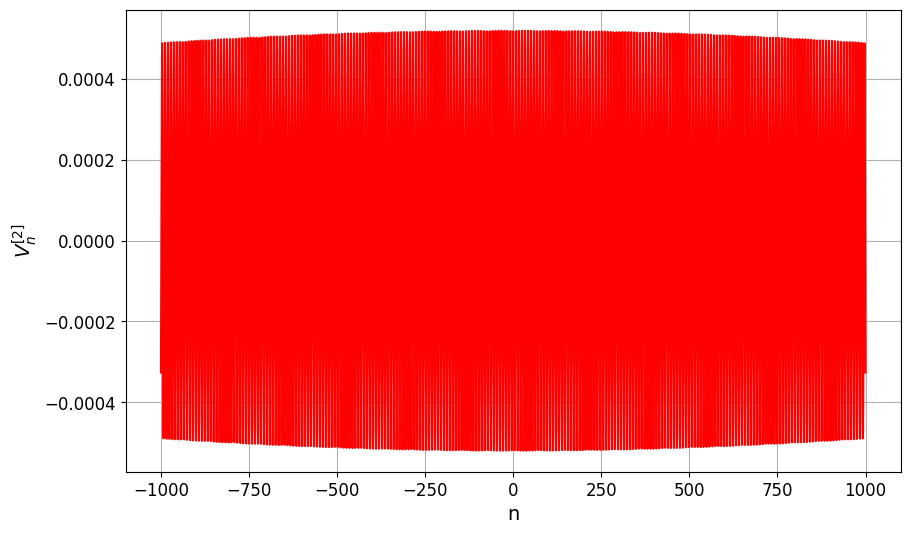}
		\end{subfigure}
		\begin{subfigure}{0.45\textwidth}
			\caption{}
			\includegraphics[width=\linewidth]{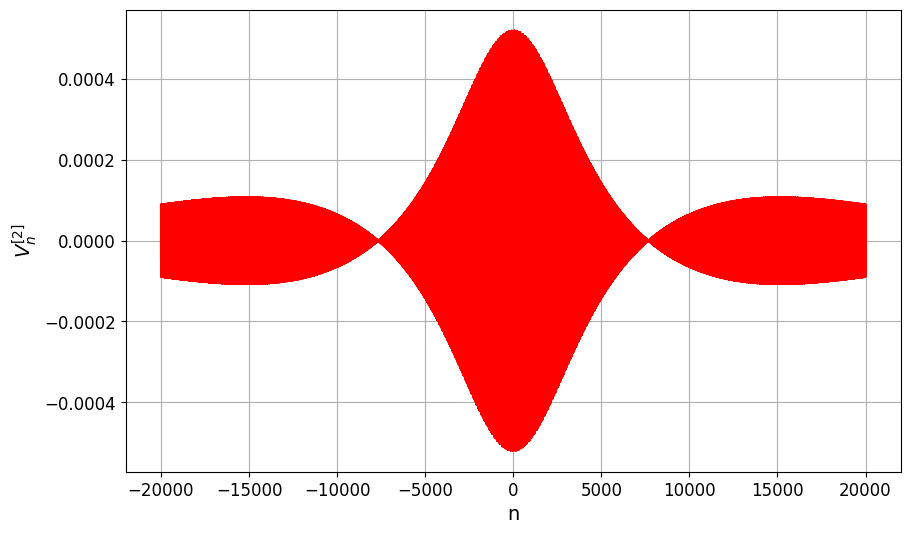}
		\end{subfigure}
	\end{center}
	\vspace{0.3cm}
	\caption{Transmission of second-order positon in an electrical transmission line (\ref{eq2}) for $\lambda=0.2+0.5i$ and different values of $\epsilon$ at $t=0$: (a),(b)  $\epsilon=0.001$ and different $n$ ranges; (c), (d) $\epsilon=0.0001$ and different values of $n$ ranges.}
	\label{fig3ep}
\end{figure*} 

In Fig. \ref{fig3ep}, the dynamics of the second-order positon in the transmission line are shown for the same parameter value, tha is $\lambda=0.2+0.5 i$ with different values of $\epsilon$ at $t=0$. The first-row figures, \ref{fig3ep}(a) and \ref{fig3ep}(b), correspond to $\epsilon=0.001$. In Fig. \ref{fig3ep}(a), the amplitude of the positon reaches approximately $0.005$, but the positon is not fully localized within the region $n=(-1000,1000)$. In contrast, Fig. \ref{fig3ep}(b) shows that the positon becomes fully localized, but over a wider range, $n=(-5000,5000)$. Similarly, in the second row, for $\epsilon=0.0001$, a similar pattern is observed. However, the amplitude decreases further, reaching approximately $0.0005$. From this, we conclude that the small parameter $\epsilon$ plays a significant role in determining both the amplitude of the second-order positon and its localization behavior. Specifically, for larger values of $\epsilon$, the second-order positon is more easily observed within a smaller range of $n$. Conversely, for smaller values of $\epsilon$, the positon's localization becomes less evident, in the lower range of $n$, and its amplitude decreases significantly. It is clear from the above analysis, that the parameter $\epsilon$ is crucial for controlling the amplitude and localization of the second-order positon in the transmission line. Larger values of $\epsilon$ result in higher amplitudes and more noticeable positon structures within smaller ranges of $n$, making them easier to observe. Conversely, smaller values of $\epsilon$ lead to reduced amplitudes and require larger ranges of $n$ for positons to fully localize. Therefore, $\epsilon$ serves as a tuning parameter to adjust the visibility and spatial localization of positons, enabling better analysis and application of their dynamics in transmission line systems.

To generate the third-order positons in the electrical transmission line described by Eq. (\ref{eq2}), we substitute Eqs. (\ref{ep11}), along with the experimental coordinates (\ref{eq11}), (\ref{eq8}), and (\ref{eq81}), into (\ref{eq3}). This gives the desired solution. We do not explicitly present the solution with experimental coordinates because it is very lengthy. Instead, we use it directly for plotting.

\begin{figure*}[!ht]
	\begin{center}
		\begin{subfigure}{0.45\textwidth}
			\caption{}
			\includegraphics[width=\linewidth]{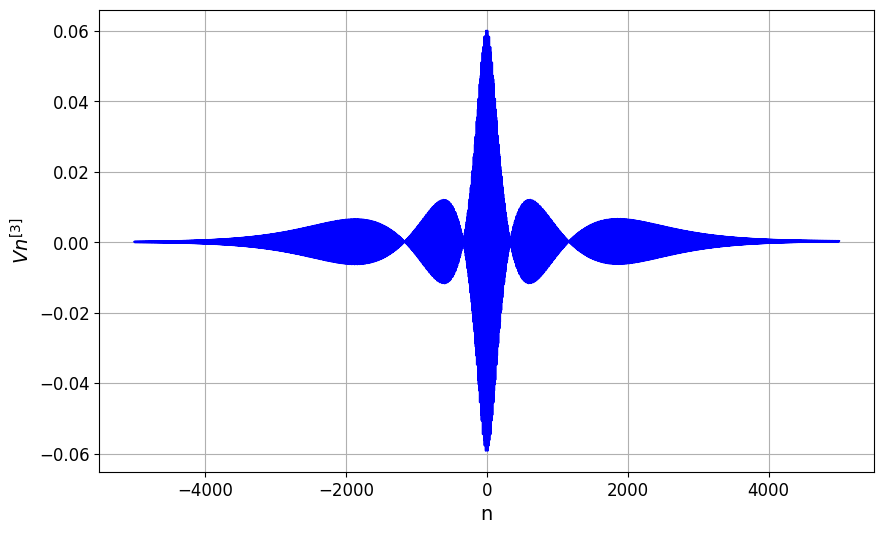}
		\end{subfigure}
		\begin{subfigure}{0.45\textwidth}
			\caption{}
			\includegraphics[width=\linewidth]{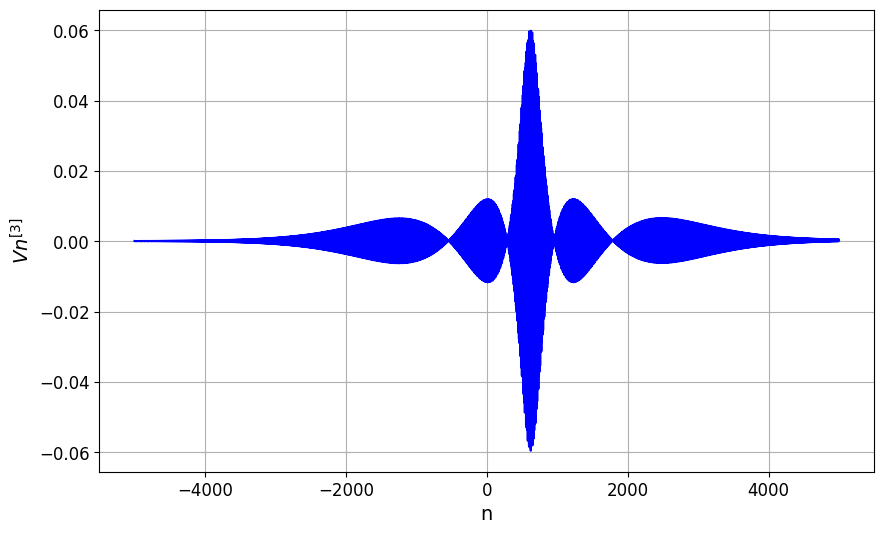}
		\end{subfigure}\\
		\begin{subfigure}{0.45\textwidth}
			\caption{}
			\includegraphics[width=\linewidth]{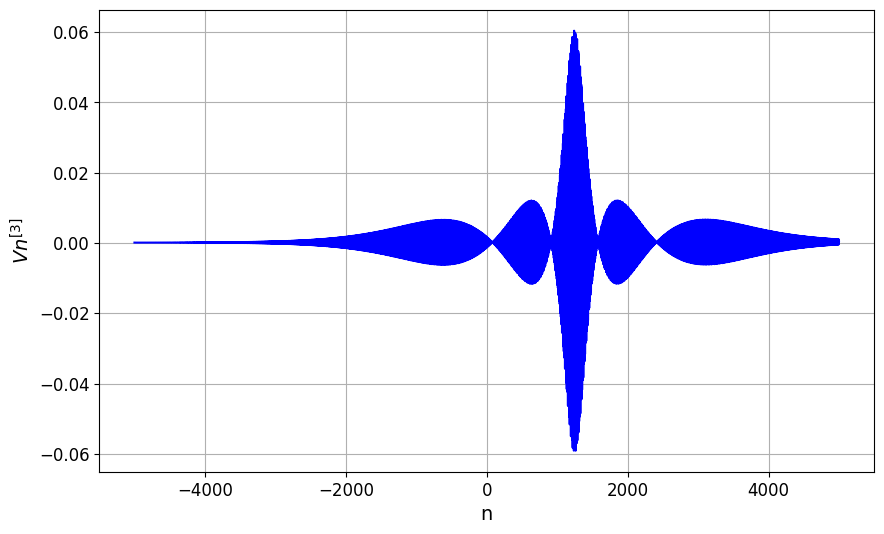}
		\end{subfigure}
		\begin{subfigure}{0.45\textwidth}
			\caption{}
			\includegraphics[width=\linewidth]{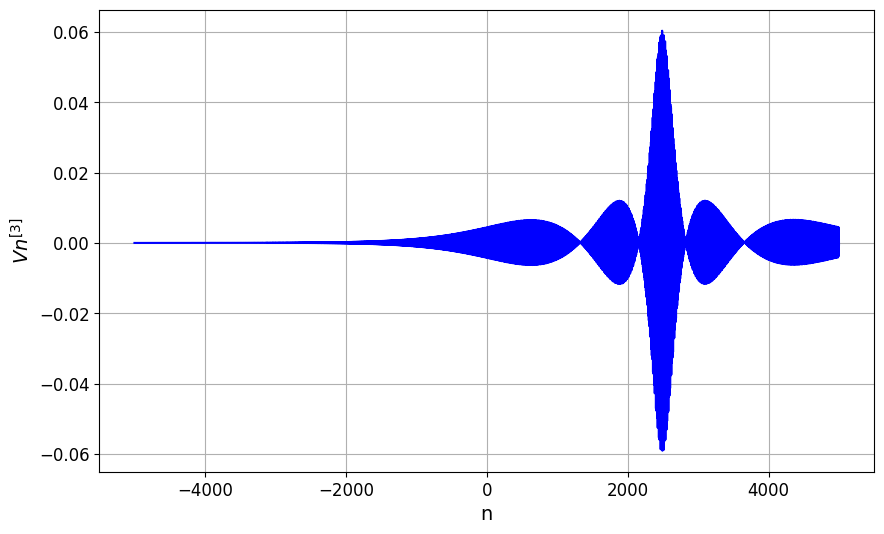}
		\end{subfigure}
	\end{center}
	\vspace{0.3cm}
	\caption{Transmission of third-order positon in an electrical transmission line (\ref{eq2}) for $\lambda=0.2+0.5i$ and different times: (a)  $t=0$; (b) $t=10$; (c) $t=20$; (d) $t=50$.}
	\label{fig5}
\end{figure*} 
Figure \ref{fig5} shows how a third order positon travels through the modified electrical transmission line described by Eq. (\ref{eq2}), using the same parameters as in Fig. \ref{fig4}. Here, we vary the time $t$ to observe how the positon moves over time. In Fig. \ref{fig5}(a), at $t=0$, the positon starts at the origin. As time increases to $t=10$ and $t=20$, the positon moves forward, as shown in Figs. \ref{fig5}(b) and \ref{fig5}(c). When the time is further increased to $t=50$, the positon reaches close to $t=4000$, where $n$ is the cell number. This demonstrates that the positon moves steadily through the network as time progresses, maintaining its shape throughout.

\begin{figure*}[!ht]
	\begin{center}
		\begin{subfigure}{0.45\textwidth}
			\caption{}
			\includegraphics[width=\linewidth]{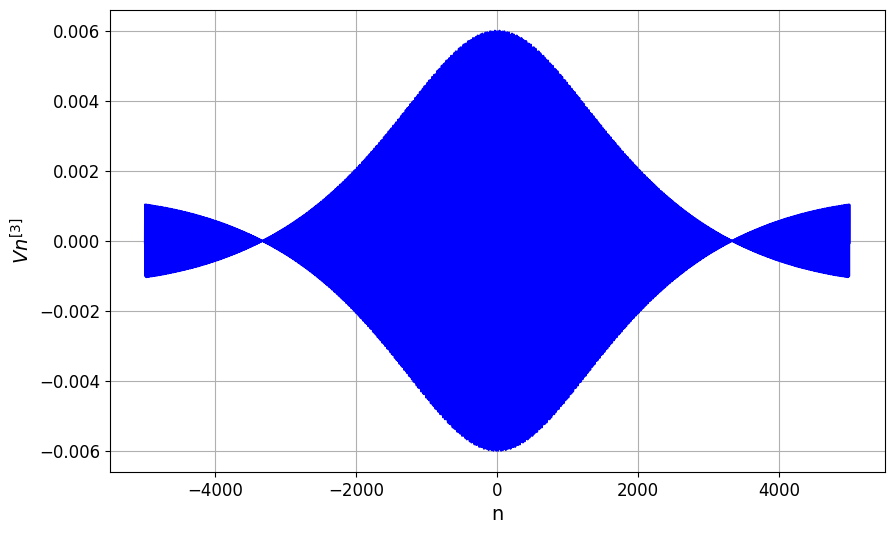}
		\end{subfigure}
		\begin{subfigure}{0.45\textwidth}
			\caption{}
			\includegraphics[width=\linewidth]{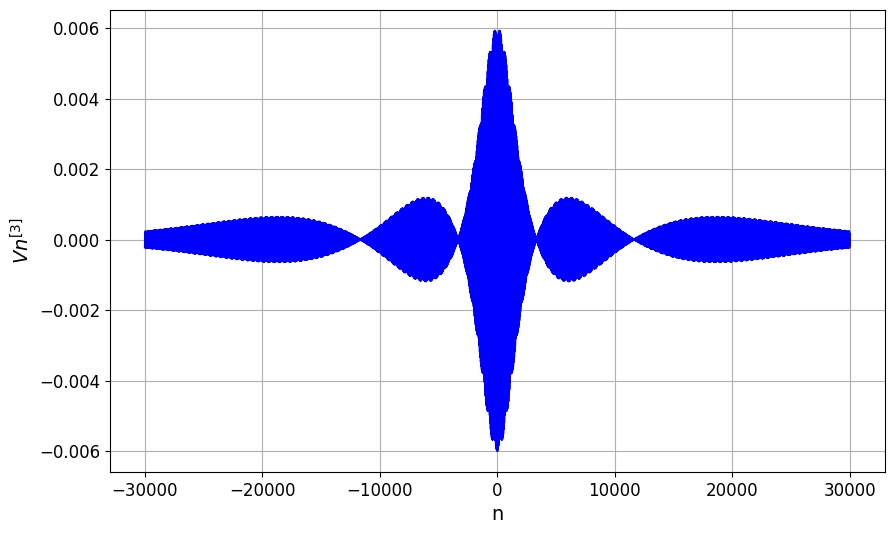}
		\end{subfigure}\\
		\begin{subfigure}{0.45\textwidth}
			\caption{}
			\includegraphics[width=\linewidth]{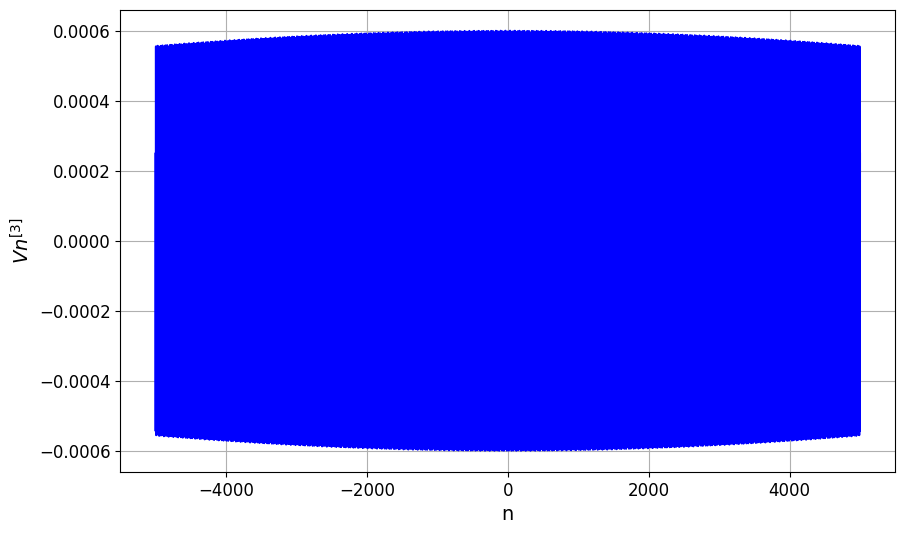}
		\end{subfigure}
		\begin{subfigure}{0.45\textwidth}
			\caption{}
			\includegraphics[width=\linewidth]{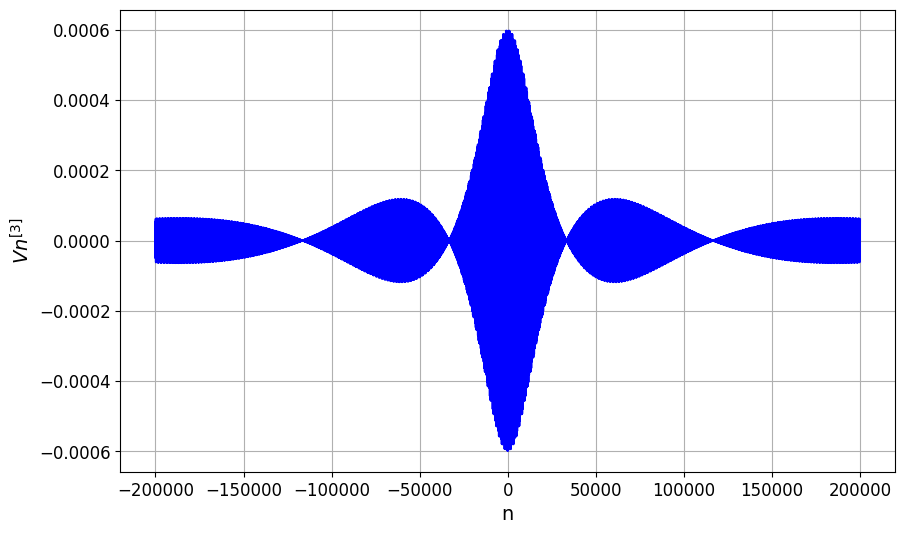}
		\end{subfigure}
	\end{center}
	\vspace{0.3cm}
	\caption{Transmission of third-order positon in an electrical transmission line (\ref{eq2}) for $\lambda=0.2+0.5i$ and different values of $\epsilon$ at $t=0$: (a),(b)  $\epsilon=0.001$ and different $n$ ranges; (c), (d) $\epsilon=0.0001$ and different values of $n$ ranges.}
	\label{fig4ep}
\end{figure*}

In Fig. \ref{fig4ep}, the dynamics of the third-order positon in the transmission line are shown for $\lambda=0.2+0.5i$ with different values of $\epsilon$ at $t=0$. The first-row figures, \ref{fig4ep}(a) and \ref{fig4ep}(b), correspond to $\epsilon=0.001$. In Fig. \ref{fig4ep}(a), the amplitude of the third-order positon reaches approximately 0.006, but the positon is not fully localized within the region $n=(-4000,4000)$. In contrast, Fig. \ref{fig4ep}(b) shows that the positon becomes fully localized, but over a wider range, $n=(-30000,30000)$. Similarly, for $\epsilon=0.0001$ (second row), a similar pattern is observed. However, the amplitude decreases further, reaching approximately $0.0006$. Now, we compare the differences between the propagation of second- and third-order positons in the transmission. The amplitude of the third-order positon is higher compared to the lower-order positon, and the width of the third-order positon is broader. Additionally, the center peaks of third-order positon have two broader humps in the background peaks. In contrast, for the second-order positon, as shown in Fig. \ref{fig3ep}, the peaks have lower amplitude and smaller width, and they have only one broader hump in the background waves.

\section{Conclusion}
In this work, we have analyzed the transmission of positons in a modified Noguchi electrical transmission line. Positons, being degenerate solitons, exhibit unique characteristics compared to other solutions. Using the reductive perturbation method, we derived the NLS equation from the circuit equations and obtained second- and third-order positon solutions through the Darboux transformation. These solutions were then applied to the modified Noguchi nonlinear transmission line circuit equation. Our analysis demonstrates how the second-order positon solution transmits within the network without altering its shape. Additionally, we explored the influence of various parameters on the positon dynamics, providing valuable insights into their behavior in experimental settings. First, by varying time, we observed that both the second- and third-order positons move, but their shape and amplitude remain unchanged. The small parameter $\epsilon$ strongly influences the amplitude and localization of the positons. Larger values of $\epsilon$ lead to more localized positons, whereas smaller values result in reduced localization and amplitude. This study establishes positons may be considered as a significant and novel one to the field of nonlinear electrical transmission line research. 
\section*{Acknowledgments}
NS wishes to thank DST-SERB, Government of India for providing National Post-Doctoral Fellowship under Grant No. PDF/2023/000619. The work of MS forms a part of a research project sponsored by Council of Scientific and Industrial Research (CSIR) under the Grant No. 03/1482/2023/EMR-II. K.M. acknowledges support from the DST-FIST Programme (Grant No. SR/FST/PSI-200/2015(C)).
\section*{Data Availability}
This manuscript has no associated data.
\section*{Conflict of interest}
The authors declare that they have no known competing financial interests or personal relationships that could have appeared to influence the work reported in this paper.

\end{document}